\documentclass[12pt]{article}

\usepackage{fancyhdr,titlesec,geometry,ragged2e}
\usepackage[dvipsnames]{xcolor}
\usepackage[many]{tcolorbox}
\usepackage{layout}
\usepackage{lipsum}
\usepackage{hyperref}
\hypersetup{
  colorlinks,
  citecolor=red,
  linkcolor=blue,
  linktoc=all
}

\usepackage[T1]{fontenc}                            
\usepackage{lmodern,mathrsfs}

\usepackage[shortlabels]{enumitem}
\usepackage{mathtools,amssymb,amsfonts,amsthm,bm,mathrsfs}   
\usepackage{array,tabularx,booktabs}                
\usepackage{graphicx,wrapfig,float,caption,epsfig}         
\usepackage{setspace,multicol}                      
\usepackage{tikz,tikz-cd,physics,cancel}            
\usepackage{esint}
\allowdisplaybreaks

\usepackage{parskip}
\DeclareGraphicsExtensions{.pdf,.png,.jpg}
\usepackage{geometry}
\renewcommand{\thefootnote}{\arabic{footnote}}

\titleformat{\section}[hang]{\normalfont\large\bfseries}{\thesection.}{3mm}{}
\titlespacing{\section}{0mm}{15mm}{4mm}

\titleformat{\subsection}[hang]{\normalfont\bfseries}{\thesubsection}{2mm}{}
\titlespacing{\subsection}{0mm}{10mm}{4mm}

\titleformat{\subsubsection}[hang]{\normalfont\bfseries}{\thesubsubsection}{5mm}{}
\titlespacing{\subsubsection}{0mm}{5mm}{1mm}

\newtheoremstyle{thmstyle}
{15pt} 
{3pt} 
{\itshape} 
{} 
{\bfseries} 
{.} 
{0.5em} 
{} 

\theoremstyle{remark}{

}

\theoremstyle{thmstyle}{

\newtheorem{lemma}{Lemma}[section]

}

\newlength{\extraspace}
\newlength{\extraspaces}
\newcommand{\be}{\begin{equation}
\addtolength{\abovedisplayskip}{\extraspaces}
\addtolength{\belowdisplayskip}{\extraspaces}
\addtolength{\abovedisplayshortskip}{\extraspace}
\addtolength{\belowdisplayshortskip}{\extraspace}}
\newcommand{\ee}{\end{equation}}
\newcommand{\ba}{\begin{eqnarray}
\addtolength{\abovedisplayskip}{\extraspaces}
\addtolength{\belowdisplayskip}{\extraspaces}
\addtolength{\abovedisplayshortskip}{\extraspace}
\addtolength{\belowdisplayshortskip}{\extraspace}}
\newcommand{\ea}{\end{eqnarray}}
\newcommand{\bas}{\begin{eqnarray*}
\addtolength{\abovedisplayskip}{\extraspaces}
\addtolength{\belowdisplayskip}{\extraspaces}
\addtolength{\abovedisplayshortskip}{\extraspace}
\addtolength{\belowdisplayshortskip}{\extraspace}}
\newcommand{\eas}{\end{eqnarray*}}

\renewcommand{\dd}{\partial}

\newcommand{\nonum}{\nonumber \\[1.5mm]}
\newcommand{\is }{&\!\!=\!\!&} 

\DeclarePairedDelimiterX\set[1]\lbrace\rbrace{#1}

\newcommand{\1}{\mbox{1\hspace{-.8ex}1}}

\newcommand{\vp}{\varphi}
\newcommand{\eps}{\epsilon}

\newcommand{\cH}{{\cal H}}

\newcommand{\cL}{{\cal L}}

\newcommand{\cO}{{\cal O}}

\newcommand{\cV}{{\cal V}}

\newcommand{\R}{\mathbb{R}}

\def\fnum@figure{\textbf{\figurename\nobreakspace\thefigure}}
\def\fnum@table{\textbf{\tablename\nobreakspace\thetable}}

\newcommand{\smallcap}[1]{{\normalfont\textsc{#1}}}

\begin{document}

\begin{titlepage}

\renewcommand{\thefootnote}{\fnsymbol{footnote}}
\makebox[1cm]{}
\vspace{1cm}

\begin{center}
  \mbox{{\Large \bf Gravitational observables}}\\[3mm]
  {{\Large \bf from a post-Carrollian expansion}}
\vspace{1.8cm}

{\sc Max Niedermaier}\footnote{email: {\tt mnie@pitt.edu}}
{\sc and} 
{\sc Mahdi Sedighi Jafari}
\\[8mm]
{\small\sl Department of Physics and Astronomy}\\
{\small\sl University of Pittsburgh, 100 Allen Hall}\\
{\small\sl Pittsburgh, PA 15260, USA}
\vspace{10mm}

\vspace{10mm} 

{\bf Abstract} \\[5mm]
\begin{quote}
 Einstein gravity, with a few exceptions, only admits {\it nonlocal} observables.
 This feature is often attributed to diffeomorphism invariance, since a gravitational
field can thereby locally be transformed away. We show this attribution to be
incorrect by constructing infinitely many exact ultralocal Dirac observables in
(velocity dominated aka electric) Carroll gravity. The latter describes
Einstein gravity's ultrarelativistic limit, is fully diffeomorphism invariant,
and of independent interest in a variety of contexts. Starting from
these Carroll observables a systematic construction of Dirac observables for
Einstein gravity is outlined. The resulting ``post-Carrollian'' expansion is organized
by the number of spatial gradients or fractional inverse powers of Newton's constant,
where each partial sum is local. 
\end{quote}  
\end{center}

\vfill

\setcounter{footnote}{0}
\end{titlepage}


{\bf Introduction.} The paucity of local observables in Einstein gravity hampers
further progress in gravitational physics by limiting options to: 
compare different approaches \cite{ASvsCDT,DonnGidd1},
isolate subsystems \cite{DonnGidd1}, or control the nonperturbative
continuum limit of discretizations \cite{Constrdisc}. 
In spatially closed geometries no local observables exist 
in vacuum on the basis of NoGo theorems \cite{ATorre1,ATorre2,Marchildon}.
This feature is thought to persist for coupling to 
generic matter, and is witnessed by perturbative computations
\cite{DonnGidd1,Froeb,GiddPerk,Cheung,DittTamb}. 
It is often attributed to diffeomorphism invariance:
since a gravitational field can locally be transformed away,
there is little to observe pointwise. The NoGo results sharpen
this to the absence of local observables, i.e.~ones built from the dynamical variables
and their derivatives at any finite order. Conversely, the
two results reported here should allow one to reframe
these traditional obstructions: in fully diffeomorphism invariant
Carroll gravity there exist infinitely many ultralocal observables.
These enter as the zeroth order of a systematic construction of
-- seemingly impossible -- local approximants to Einstein observables. 

Carroll gravities describe Einstein gravity's
ultrarelativistic limit and together with their geometry are an
active area of current research \cite{Primer,CarrollHolo,CarrollHoloRept,CarrollGeom}.
There are several variants of which we focus here on  
the original, velocity dominated (aka electric) limit as introduced
by Isham \cite{Isham} and Henneaux \cite{Henn1}; see \cite{Carrollgrav1}
for the terminology and \cite{Hartong,Carrollgrav2} for a first order
formulation.
For short, we shall refer to it as Velocity Dominated Carroll (VDC) gravity.
For our purposes its
second order Hamiltonian formulation is advantageous. The action
minimally coupled to a scalar field reads 
\begin{eqnarray}
\label{VDCaction} 
S \!&=& \!\int_{t_i}^{t_f}\!dt \int_{\Sigma} \!d^3x \, 
\big\{ p^{ab} \partial_t q_{ab} + p^{\varphi} \partial_t \varphi
-\nu^a {\cal H}_a - n {\cal H} \big\}\,,
\nonumber
\\
{\cal H}_a \!&=&\! -2 \nabla_{\!b} p_a^{\;b} + p^{\varphi} \partial_a \varphi\,,
\\ {\cal H} \!&=&\! 2 \kappa \Big( p_a^{\;b} p_b^{\;a} - \frac{1}{2}p^2
+ \frac{1}{4} (p^{\varphi})^2\Big) + \frac{1}{\kappa} q(U(\varphi) + \Lambda)\,.
\nonumber
\end{eqnarray}
Here ${\cal H}_a, a=1,2,3$, is the diffeomorphism constraint and ${\cal H}$
is the Hamiltonian constraint. Both are built from the canonical pairs
$(p^{ab}, q_{ab})$, $(p^{\varphi}, \varphi)$, which obey standard
Poisson brackets. Indices are moved with $q_{ab}$, $\nabla_{\!a}$ is 
the associated covariant derivative, and $q=\det(q_{ab}),\, p= p_a^a$. 
The scalar field $\varphi$ has been
rendered dimensionless by the rescaling $\varphi \mapsto
\sqrt{2 \kappa} \varphi$ and is selfinteracting via the potential
$U(\varphi)$. The Lagrange multipliers $(n,\nu)$ play the role of
lapse and shift, and $\kappa,\Lambda$ are the Newton and
cosmological constants. Finally, the underlying manifold $M =
\R \times \Sigma$ is foliated in terms of spacelike hypersurfaces,
which we assume to be compact and closed, for simplicity.
The action (\ref{VDCaction}) differs from
its counterpart $S_{\smallcap{e}}$ for Einstein gravity merely by the absence of
a term $q(-R(q) + q^{ab} \partial_a \varphi \partial_b \varphi)/(2 \kappa)$
in the Hamiltonian constraint, where $R(q)$ is the Ricci scalar built
from the spatial metric $q_{ab}$. The action $S$ can be obtained from
$S_{\smallcap{e}}$ by $S= \lim_{\ell \rightarrow \infty} S_{\smallcap{e}}|_{\ell}$,
where $S_{\smallcap{e}}|_{\ell}$ has spacelike distances enhanced via
$q_{ab}|_{\ell} = \ell^2 q_{ab}$, $p^{ab}|_{\ell} = \ell^{-2} p^{ab}$, $\ell \gg 1$,
combined with $\kappa |_{\ell} = \ell^3 \kappa$ \cite{SvsCtensor}. Since
$\kappa \propto G_{\smallcap{n}}/c^3$, this can alternatively be viewed
as a limit of large Newton constant $G_{\smallcap{n}}$ or small
speed of light $c$.%
\footnote{ The commonly used direct decomposition in powers of $c$ 
  renders scalar selfinteractions subleading and thus does not
  produce Eq.~(\ref{VDCaction}).}

{\bf Dirac observables in VDC gravity.} 
Given the textbook-advertised uniqueness properties of the Einstein-Hilbert
action (from which the Hamiltonian version $S_{\smallcap{e}}$ derives)
it is surprising that (\ref{VDCaction}) is also fully diffeomorphism invariant.  
This is implicit in \cite{Henn1} and can be understood through the
redundancy \cite{SvsCtensor} of the so-called Ehresmann connection.
Concretely, one has $\delta_{\epsilon}^{\smallcap{h}} S =0$ \cite{GRstrongI},
where the Hamiltonian gauge variations $\delta_{\epsilon}^{\smallcap{h}} X$,
$X = (n, \nu^a), (q_{ab}, p^{ab}), (\varphi, p^{\varphi})$, are
parameterized by descriptors $\epsilon(t,x), \epsilon^a(t,x)$ that
are in one-to-one correspondence to linearized diffeomorphisms
along the standard lines \cite{Pons}. To match $n$ we take the temporal
descriptor $\epsilon$ to be
a spatial $-1$ density that vanishes at $t = t_i,t_f$. The gauge variations
other than $\delta^{\smallcap{h}}_{\epsilon} n, \delta^{\smallcap{h}}_{\epsilon} \nu^a$
are canonically generated via $\delta_{\epsilon}^{\smallcap{h}} F = \{ F, \int\! d^3x[
  {\cal H} \epsilon + {\cal H}_a \epsilon^a]\}$.    
Related to its full diffeomorphism invariance the action (\ref{VDCaction})
also propagates the same number of physical degrees of freedom as
its Einstein counterpart.

Dirac observables for (\ref{VDCaction}) are essentially gauge invariant quantities,
$\delta^{\smallcap{h}}_{\epsilon} {\cal O}\approx 0$, where `$\approx$' means
equality modulo ${\cal H}_a =0 = {\cal H}$. For the solutions
we find, the vanishing holds exactly; moreover they are built from
$(q_{ab}, p^{ab}), (\varphi, p^{\varphi})$ only.  Our Dirac observables for
VDC gravity therefore satisfy  
\begin{equation}
\label{Diracobs}   
\{ {\cal H}_a, {\cal O}\} = 0 =
\{ {\cal H}, {\cal O}\} \,,
\end{equation}
which is a strong version of the standard characterization \cite{HTbook}. 
Without further ado we introduce now the building blocks 
for the construction of such ${\cal O}$'s:
\begin{eqnarray}
\label{bblocks} 
\pi_{\!a}^{\;b} &:=& q_{ac}(p^{cb} - \frac{1}{3} q^{cb} p)\,,
\nonumber
\\
E &:=& \frac{\kappa}{3} p^2 - \frac{\kappa}{2} (p^{\varphi})^2
- \frac{1}{\kappa} q ( U(\varphi) + \Lambda)\,.
\end{eqnarray}
These are spatial densities of weight $+1$, $+2$, respectively. 
Further ${\cal H} = 2 \kappa p_a^{\;b} p_b^{\;a} - (2 \kappa p^2/3 + E)$, 
so that we take $E \geq 0$ in the following (in case one would want to
solve ${\cal H} =0$, which will not be done here). The interest
in the quantities (\ref{bblocks}) lies in the fact that they are
invariant under temporal gauge transformations:
$\delta_{\epsilon}^{\smallcap{h}}\pi_{\!a}^{\;b}= {\cal L}_{\vec{\epsilon}}
\pi_{\!a}^{\;b}$,
$\delta_{\epsilon}^{\smallcap{h}} E= {\cal L}_{\vec{\epsilon}} E$, where
${\cal L}_{\vec{\epsilon}}$ is the spatial Lie derivative, acting according to the
tensor density type under consideration. This holds off-shell, i.e.~without
using any equations of motion. For a scalar $+1$ density $F$ one has in
particular ${\cal L}_{\vec{\epsilon}} F = \partial_a( \epsilon^a F)$, so that
its spatial integral over a closed spacelike hypersurface $\Sigma$
vanishes. In addition to $E$ there are two other building blocks
$Q_1, P_1$ locally built from $q,p, \varphi, p^{\varphi}$. These also obey  
$\delta_{\epsilon}^H Q_1 = {\cal L}_{\vec{\epsilon}} Q_1$,
$\delta_{\epsilon}^H P_1 = {\cal L}_{\vec{\epsilon}} P_1$, and form a
canonical pair. However, since for nontrivial potential $U(\varphi)$
no explicit expressions are available for them, we refrain from using them
here directly. 

The strategy in the following is simply to form nonlinear combinations
of (\ref{bblocks}) that are spatial $+1$ densities $F$. These combinations
will still be invariant under temporal gauge variations but obey
$\delta^{\smallcap{h}}_{\epsilon} F = \partial_a( \epsilon^a F)$. Aiming at such
combinations we start with the traces of the matrix powers of $\pi_{\!a}^{\;b}$.
By the Cayley-Hamilton theorem $\pi^3 - ({\rm tr} \pi^2) \pi/2 - \1
\det \pi =0$ holds; so one may take ${\rm tr} \pi^2$ and ${\rm tr}\pi^3$
as independent traces.
Augmenting $E$, one can build ratios of homogeneous polynomials in
the three density $+1$ variables $({\rm tr} \pi^2)^{1/2},
({\rm tr} \pi^3)^{1/3}$, $E^{1/2}$. Any ratio of total density weight $+1$ 
qualifies as an $F$ with the above properties. For illustration, we note
as simple examples 
\begin{equation}
\label{Fexamples}
F = \frac{ ({\rm tr} \pi^3)^{n/3}}{ ({\rm tr} \pi^2)^{\frac{n-1}{2}} }\,,
\quad F = \frac{E^{n/2}}{({\rm tr} \pi^3)^{\frac{n-1}{3}}}\,, \quad n \in
\mathbb{N}\,.
\end{equation} 

\noindent
{\bf Result}: Velocity dominated Carroll gravity admits infinitely
many strong, off-shell Dirac observables  of the form
${\cal O}_F = \int_{\Sigma}\,d^3 x F$, where $F$ is a rational function
of net spatial density weight $+1$ in 
$({\rm tr}\pi^2)^{1/2}, ({\rm tr}\pi^3)^{1/3}, E^{1/2}$.
\medskip

\noindent
These observables exist both in vacuum (where $E^{1/2} \propto  |p|$) and
upon minimal coupling to a scalar field $\varphi$. They are ultralocal  
in the sense that $F$ is built from the phase space
variables $(q_{ab}, p^{ab})$, $(\varphi, p^{\varphi})$ at a
spacetime point $(t,x)$. Additional expressions like (\ref{Fexamples})
could be formed using the before-mentioned $Q_1,P_1$ building blocks. 
As a step towards a general NoGo theorem
Kuchar \cite{KucharNoGo} performed a systematic 
search for Dirac observables in vacuum VDC gravity that are linear
in the momenta, and found only $\int\! d^3 x \,p$. This
is consistent
with the above result in that all other admissible $F$'s are nonlinear.  

{\bf Additional symmetries and evasion of NoGo results.} 
From the viewpoint of Einstein gravity the existence of these ultralocal
Dirac observables is puzzling because there such a quantity cannot exist
by the NoGo theorems \cite{TorreNoGo,ATorre1,ATorre2,Marchildon}.
Somehow VDC gravity must evade the NoGo results, and we are led to inquire:
how? It cannot be blamed on a breaking of diffeomorphism invariance
because, as stressed before, VDC gravity is still fully diffeomorphism
invariant. The key step in establishing the NoGo results
is to classify all so-called generalized symmetries, i.e.~operations
that map a solution of the linearized field equations into another
one. The result is \cite{ATorre1,ATorre2,Marchildon} that all such
operations are produced either by linearized diffeomorphisms or by
rescalings. There are no hidden symmetries. Though initially shown
in a Lagrangian formulation, the result can be transferred to phase
space and then entails the absence of local Dirac observables
in Einstein gravity \cite{TorreNoGo}. In view of this line of reasoning
one is led to suspect that VDC gravity evades the NoGo results by
having hidden symmetries. Indeed, this turns out to be the case.
Moreover, these additional symmetries are off-shell symmetries of the nonlinear action, not just symmetries of the
linearized field equations. 

To formulate the result, let $F$ be an ultralocal solution of $\{ F,{\cal H}\} =0$.
Explicitly, $F = F(q_{ab},p^{ab},\varphi, p^{\varphi})$ is a function of phase
space variables but not their derivatives that solves
\begin{equation}
\label{FPDE}
  \frac{\partial {\cal H}}{\partial p^{ab}}
  \frac{\partial F}{\partial q_{ab}} 
-  \frac{\partial {\cal H}}{\partial q_{ab}}
  \frac{\partial F}{\partial p^{ab}} 
+  \frac{\partial {\cal H}}{\partial p^{\varphi}}
  \frac{\partial F}{\partial \varphi} 
-  \frac{\partial {\cal H}}{\partial \varphi}
  \frac{\partial F}{\partial p^{\varphi}} =0. 
\end{equation} 
If $F$ is also a spatial $+1$ density we introduce local gauge
variations by 
\begin{eqnarray}
\label{Fvars}   
\delta^{\smallcap{f}} q_{ab} &=& \frac{\partial F}{\partial p^{ab}}\,,\;\;
\delta^{\smallcap{f}} p^{ab} = -\frac{\partial F}{\partial q_{ab}}\,,
\nonumber \\
\delta^{\smallcap{f}} \varphi &=& \frac{\partial F}{\partial p^{\varphi}}\,,\;\;
\delta^{\smallcap{f}} p^{\varphi} = -\frac{\partial F}{\partial \varphi}\,, 
\end{eqnarray} 
and $\delta^{\smallcap{f}} n =0$, $\delta^{\smallcap{f}}\nu^a =0$.
Then $\delta^{\smallcap{f}}$ is a local symmetry of VDC gravity in the sense that
\begin{equation}
\label{SFinv}   
\delta^{\smallcap{f}} S = \int_{\Sigma} \! d^3 x
\Big( p^{ab} \frac{\partial F}{\partial p^{ab}} +
p^{\varphi} \frac{\partial F}{\partial p^{\varphi}} -F \Big)\bigg|_{t_i}^{t_f}\,.
\end{equation} 
Since the action is preserved up to a boundary term the gauge variations
preserve the Hamiltonian equations of motion. For the constraints
${\cal H}$, ${\cal H}_a$ this holds exactly, while for the evolution
equations  it holds to linear order in $\delta^{\smallcap{f}}$. Importantly,
these symmetries are unrelated to diffeomorphisms and we shall refer 
to them as ``additional local symmetries'' of VDC gravity. 
A further analysis shows that they also give rise to
generalized symmetries in the sense of \cite{ATorre1,ATorre2,Marchildon},
which thereby explains how the NoGo results for Einstein gravity
are evaded in VDC gravity.

Of course (\ref{SFinv}) is only relevant if ultralocal solutions of
(\ref{FPDE}) exist. But just such solutions have been found in the
earlier result, for example (\ref{Fexamples}). Ultralocal Dirac
observables for VDC gravity therefore give rise to additional local
symmetries and vice versa. The reason for
spelling out (\ref{FPDE}) is that it highlights the key difference to
Einstein gravity. Since ${\cal H}$ is ultralocal in VDC gravity,
the equation (\ref{FPDE}) is for each spacetime point $(t,x)$
a linear first order partial differential equation. As such it
is soluble by  the method of characteristics. The characteristics are
governed by a system of nonlinear ordinary differential equations,
which turn out \cite{Trivgrad} to coincide, pointwise in $x$, 
with the evolution equations for Bianchi I spacetimes in general relativity
\cite{Ellisbook}. The pointwise counterpart
of (\ref{FPDE}) thereby defines conservation laws for the Bianchi I
evolution. From this viewpoint, one can recover the
previously found $F$'s via a suitable lift (functions of $t$ to functions
of $(t,x)$) of Bianchi I conservation laws.  

{\bf From Carroll to Einstein.} 
So far, we treated VDC gravity as a gravity theory in its own right with
properties intriguingly different from Einstein gravity.
One can alternatively regard Carroll gravity as the leading term
in a suitable decomposition of an action principle for
Einstein gravity. With different initial actions such decompositions
have been presented in
\cite{Carrollexp1,SvsCtensor,Carrollgrav1}.
These can be used as the starting point for the construction or
expansion of solutions in a tradition dating back to the
analysis of Belinski-Khalatnikov-Lifshitz, see
\cite{Carrollexp1,RingstroemCR}.  
In the present context, this raises the possibility of constructing
observables for Einstein gravity in a new type of expansion. 
For the standard perturbative expansion the computation of observables for Einstein gravity
has been pursued by several teams in different (also quantum) settings
and encompasses weak field, small Newton constant, small clock time
deviation, see e.g.~\cite{GiddPerk,Cheung,Froeb,DittTamb}. 
In view of the above results one is led to search for a complementary 
expansion of gravitational observables, where VDC gravity and its
simple ultralocal observables are taken as the zeroth order. If well-posed, 
such an expansion could be viewed alternatively as one for large
Newton constant or small speed of light
(``post-Carrollian''), see after Eq.~(\ref{VDCaction}). It should complement resummation
attempts for the weak coupling series \cite{PNresum1,PNresum2} and shed
light on earlier notions of quasilocality \cite{KhavkineObs,Hoehn}. In a cosmological context it is an expansion
away from the Big Bang. As a proof of principle a well-posed
expansion of this type has been obtained for Gowdy cosmologies using integrable
systems methodology \cite{Gowdyobs}.

In a Hamiltonian framework one will search for Dirac observables
in Einstein gravity in the form ${\cal O}_{\smallcap{e}} =
\int\! d^3 x F_{\smallcap{e}}$, with $F_{\smallcap{e}}$ 
a spatial $+1$ density. Then only weak commutativity
with the Hamiltonian constraint needs to be imposed,
i.e.~$\{{\cal H}_{\smallcap{e}}, {\cal O}_{\smallcap{e}} \}
|_{{\cal H}_{\smallcap{e}} = {\cal H}_a =0}=0$. Performing the scaling 
mentioned after Eq.~(\ref{VDCaction}) one has ${\ell^{-3}\cal H}_{\smallcap{e}}|_{\ell}
={\cal H} + \check{\kappa} {\cal V}$, where 
\begin{equation}
 \label{gradpot}  
{\cal V} = \frac{q}{2\kappa}\big(\!-R(q) + q^{ab} \partial_a \varphi \partial_b \varphi \big)\,,
\end{equation} 
is the gradient potential, and $\check{\kappa} = \ell^{-2}$ can
also be read as $(\bar{\kappa}/\kappa)^{2/3} \propto
\kappa^{-2/3} \propto c^2 G_{\smallcap{n}}^{-2/3}$ if
$\bar{\kappa} = \kappa \ell^{-3}$ is held fixed.
An expansion in powers of $\check{\kappa}$ can therefore be viewed
alternatively as one in the number of spatial gradients, in powers of $c^2$, or in
powers of $G_{\smallcap{n}}^{-2/3}$.  Taking one
of the ultralocal Dirac observables ${\cal O}$ of VDC gravity as the
zeroth order, ${\cal O}_0 = {\cal O}$, we consider an ansatz
of the form ${\cal O}_{\smallcap{e}} = \sum_{j \geq 0}
\check{\kappa}^j {\cal O}_j$. It is convenient to average
${\cal H}_{\smallcap{e}}$ with the descriptor $\epsilon$ of a temporal gauge
variation. The defining equation for  a Dirac observable for Einstein gravity then
reads 
$\{{\cal H}_{\smallcap{e}}(\epsilon), {\cal O}_{\smallcap{e}} \} =
{\cal H}_{\smallcap{e}}(A(\epsilon)) + \vec{{\cal H}}(\vec{B}(\epsilon))$, 
for some coefficient functions $A(\epsilon)$, $\vec{B}(\epsilon)$ that
may depend on the phase space variables. In the present framework  
these will be expanded according to  
$A(\epsilon) = \sum_{j \geq 0} \check{\kappa}^j A_j(\epsilon)$,
$\vec{B}(\epsilon) = \sum_{j \geq 0} \check{\kappa}^j \vec{B}_j(\epsilon)$,
where each term may depend on the VDC phase space variables and their
spatial derivatives. Inserted into the above defining equation, the lowest order
reads $\{ {\cal H}(\epsilon), {\cal O}_0\} = {\cal H}(A_0(\epsilon))
+ \vec{{\cal H}}(\vec{B}_0(\epsilon))$, and the strong vanishing of the
Poisson bracket for the VDC Dirac observables found earlier amounts to
$A_0(\epsilon) =0$, $\vec{B}_0(\epsilon) =0$. For the subleading terms
one is led to a system of recursive equations
\begin{eqnarray} 
\label{obsrecursion}
\{ {\cal H}(\epsilon), {\cal O}_j\} &=& - \{ {\cal V}(\epsilon),{\cal O}_{j-1} \} +
V(A_{j-1}(\epsilon))
\nonumber
\\ 
&+& {\cal H}(A_j(\epsilon))
+ \vec{{\cal H}}(\vec{B}_j(\epsilon)) \,,
\quad j \geq 1\,.
\end{eqnarray}
At each order, the basic structure of (\ref{obsrecursion}) is that of a so-called
``homological equation'', $\{{\cal H}(\epsilon), X\} = Y$, with $Y$ known.   
For perturbations around a finite dimensional integrable system several
solution techniques are available, see \cite{Nikolaev} and the references therein. 
These techniques do, however, not readily carry over to VDC gravity.

{\bf Solution of the recursion.} Instead, we developed a custom technique
(partly drawing on \cite{Trivgrad}) that invokes the following ingredients:
(a) A spatial $-1$ density $T$ called a ``clock field'', characterized by
$\{ T, {\cal H}(\epsilon)\} =\epsilon$.
(b) The solution of a single set of ordinary differential equations, the previously
encountered characteristics of (\ref{FPDE}). (c) An explicit recursive 
formula for the coefficients $A_j(\epsilon), \vec{B}_j(\epsilon)$, $j \geq 1$.
(d) A master solution formula relating ${\cal O}_j$ to ${\cal O}_{j-1}$.  
We now first comment consecutively on each of these ingredients and then present
the solution of  (\ref{obsrecursion}) in terms of them.

(a) There are several ways to construct a clock field. One that works
both in vacuum and for a generic interaction potential $U(\varphi)$ is
as follows. Assume that the mixed tensor $\pi_a^{\;b}$ from 
(\ref{bblocks}) is diagonalizable over the reals. Let $m_{ab} := q^{-1/3} q_{ab}$ be
the unimodular metric canonically conjugate to $\pi^{ab} = m^{ac} \pi_{\!c}^{\;b}$.
Then 
\begin{equation}
\label{clockdef}
T = \frac{{\rm tr}[ \pi {\rm Log}(m \overline{m}^{-1})]}{ 4 \kappa {\rm tr}\pi^2}\,, 
\end{equation}
is a spatial $-1$ density and satisfies $\{ T, {\cal H}(\epsilon)\} =\epsilon$.
Here $\overline{m}_{ab}$ is separately invariant, $\{\overline{m}_{ab},
{\cal H}(\epsilon)\} =0$, and built from the components of $\pi_a^{\;b}$. 
The logarithm is that for symmetric $3\times 3$ matrices. 
$T$ is a variant of a clock field in the sense of  \cite{DittTamb}, but plays
a different role here.

(b) The next ingredient is the characteristic flow of the Hamiltonian
vector field associated with ${\cal H}$. Writing $z = (q_{ab}, \varphi, p^{ab},
p^{\varphi})$ for a generic phase space point, the flow consists of
solutions $z^f(z;\tau)$ 
of the evolution equations $\partial_{\tau} z^f = \{ z^f , {\cal H} \}$,
$z^f(z;0) =z$, pointwise in $x \in \Sigma$, with  respect to a (unphysical) flow
parameter $\tau = \tau(x)$ that is a spatial $-1$ density. Backpropagating a generic
point $z$ with $T$ gives rise to coordinates $\bar{z}:=z^f(z,-T)$ that are invariant
under temporal gauge transformations, $\{ \bar{z}, {\cal H}\} =0$. 
These we use to define an interpolating  flow $z^{(s)} = z^f(z, (s-1) T(z))$, $0 \leq s \leq 1$,
that connects the original phase space point $z$ at $s=1$ to its gauge invariant
counterpart $\bar{z}$ at $s=0$. We write $z^{(s)}
= (q_{ab}^{(s)}, \varphi^{(s)}, p_{(s)}^{ab}, p_{(s)}^{\varphi})$ for the components and
use $F|_s$ to denote evaluation of a phase space function $F$
on $z^{(s)}$. 

(c) For $j\!=\!1$ the coefficients entering the solution of (\ref{obsrecursion}) read
\begin{eqnarray}
\label{AB1} 
  A_1(\epsilon) &=& \int_0^1 \! ds s (1\!-\!s) {\cal L}_{\{\vec{\gamma}(\alpha, \epsilon),
      {\cal O}_0\}|_{s, \alpha =T}} T\,,
  \nonumber
  \\
  \vec{B}_1(\epsilon) & = & - \int_0^1 \! ds s \{\vec{\gamma}(\alpha, \epsilon),
      {\cal O}_0\}|_{s, \alpha =T}\,,
 \end{eqnarray} 
where $\vec{\gamma}(\alpha, \epsilon) := q q^{ab}( \alpha \partial_b \epsilon -
\epsilon \partial_b \alpha)$, and ${\cal L}_{\vec{\xi}} T = \xi^a \partial_a T
- \partial_a \xi^a T$ is the Lie derivative of a spatial $-1$ density.  A recursive formula,
not needed here, then determines all other $A_j(\epsilon), \vec{B}_j(\epsilon)$, $j \geq 2$.

(d) The master formula relating the different orders of the expansion has 
the concise form 
\begin{equation}
\label{obsrecursionsol}
  {\cal O}_j = \int_0^1 \! ds \{ V(\alpha), {\cal O}_{j-1} \} \big|_{s, \alpha =T}\,,
  \quad j \geq 1\,.
  \end{equation} 
Technically, the key assertion is that for all $j \geq 1$ there exist
spatial $-1$ densities $A_j(\epsilon)$ and vectors $\vec{B}_j(\epsilon)$
(determined by (c)), such that (\ref{obsrecursionsol}) solves (\ref{obsrecursion}).
Since $V$ contains two spatial derivatives, it entails that the ${\cal O}_j$ are
{\it local} quantities. Specifically they are of the form
\begin{equation} 
  \label{Ojloc}
 {\cal O}_j = \int\! d^3 x \, G_j(x)\, ,
\end{equation} 
where  $G_j$ is a differential polynomial in the phase space variables
(and a spatial $+1$ density) containing at most $2j$ spatial derivatives.

{\bf Result.} Let $T$ be a clock field as in (a), let $z^{(s)}$ be the interpolating
flow from (b), and take the ${\cal O}_j$ as computed from (d).
Then, the partial sums of the series
$\sum_{j \geq 0}\check{\kappa}^j {\cal O}_j$
provide systematic {\it local} approximants to Dirac observables ${\cal O}_{\smallcap{e}}$
in Einstein gravity. The leading-order term ${\cal O}_0$ is one of the ultralocal
Dirac observables for VDC gravity described earlier. The corrections are
organized by the number of spatial gradients or by powers of
$\check{\kappa} \propto c^2 G_{\smallcap{n}}^{-2/3}$.

This is very different from the usual weak coupling (in powers of $G_{\smallcap{n}}^{1/2}$)
or post-Newtonian (in powers of $1/c$) expansions, where  all orders
are highly nonlocal. The result initiates a new approach to the intrinsic physics
of diffeomorphism invariant systems. It may be hoped that  the obstructions
mentioned in the introductory paragraph can be alleviated thereby.

\appendix

\section{The first post-Carrollian correction} 

Here we derive the formulas providing the first post-Carrollian correction
for Dirac observables. Explicitly, we claim that given an ultralocal VDC Dirac
observable $\cO_0$, and a clock field $T$, the expression
\be
\label{O1} 
\cO_1 = \int_0^1 \! ds \, \{ \cV(\alpha), \cO_0\}\big|_{s,\alpha=T} \,,
\ee
solves $\{ {\cal H}(\epsilon), {\cal O}_1\} = - \{ {\cal V}(\epsilon),{\cal O}_0 \}
+ {\cal H}(A_1(\epsilon)) + \vec{{\cal H}}(\vec{B}_1(\epsilon))$, with
$A_1(\epsilon)$, $\vec{B}_1(\epsilon)$ from (\ref{AB1}).  

We prepare the following

\begin{lemma} Let $F$ be a spatial differential polynomial in the basic
  phase space variables and a spatial $+2$ density. Write $F(\alpha) =
  \int \! d^3 x\, F(x) \alpha(x)$ for a spatial $-1$ density $\alpha$. For a
  given clock field $T$ define  $\cO_F := \int_0^1 \! ds F|_s (T)$. Then
  $\cO_F$ satisfies
  \be
  \label{OFvar} 
  \delta_{\eps,0}^H \cO_F = F(\eps) + \int_0^1 \! ds s \big[
    \delta_{\eps,0}^H F (\alpha) -
    \delta_{\alpha,0}^H F (\eps) \big] \big|_{s, \alpha =T}\,.
    \ee
In particular, $[\delta_{\eps,0}^H \cO_F - F(\eps)]|_{\eps = T} =0$. 
\end{lemma}   

{\it Proof.} The derivation is based on the interplay between the gauge
variation and the s derivatives of flowed quantities. From the definitions
of the interpolating flow $z^{(s)} = z^f(z, (s-1) T(z))$, $0 \leq s \leq 1$,
with $\dd_{\tau} z^f = \{ z^f, \cH\}$, $z^f(z;0) = z$,  one obtains the relations
\be
\label{oaux1} 
     s [\delta_{\eps,0}^{\smallcap{h}} F(\alpha)]\big|_s = 
      \delta^{\smallcap{h}}_{\eps,0} \big[F(\alpha)|_s\big]\,, \quad
      \frac{d}{ d s} \big[F(\eps)\big|_s\big] = 
           \big[ \delta^{\smallcap{h}}_{\alpha,0} F(\eps)\big]\big|_{s,\alpha=T}\,.
\ee
Then
\ba
\label{oaux2}
\delta_{\eps,0}^{\smallcap{h}} \cO_F \is  \int_0^1 \! ds \,
\Big( \!F(\eps) \big|_s+ \delta_{\eps,0}^{\smallcap{h}}[F(\alpha)|_s]\big|_{\alpha =T} \Big)  
\nonum
\is \int_0^1 \! ds \,
\Big( \!F(\eps) \big|_s+ s [\delta_{\alpha,0}^{\smallcap{h}} F(\eps)]\big|_{s,\alpha =T} 
+ s \big[\delta_{\eps,0}^{\smallcap{h}} F(\alpha)
  - \delta_{\alpha,0}^{\smallcap{h}}F(\eps)\big]\big|_{s,\alpha =T} \Big) 
\nonum
\is \int_0^1 \! ds \, \frac{d}{ds} \big(s F(\eps)|_s\big) + 
\int_0^1 \! ds \, s
\big[\delta_{\eps,0}^{\smallcap{h}} F(\alpha)
  - \delta_{\alpha,0}^{\smallcap{h}}F(\eps)\big]\big|_{s,\alpha =T} \,.
\ea
The first term evaluates to $sF(\eps)|_{s=1} - s F(\eps)|_{s=0}= F(\eps)$,
where $z^{(s=1)} = z$ (the original phase space point) and $z^{(s=0)} = \bar{z}$
(its $\delta^{\smallcap{h}}_{\eps,0}$ invariant counterpart) enter. \qed 

One application of the Lemma is to the $j\!=\!1$ instance 
of (\ref{obsrecursion}). Since $A_0(\eps) =0$ the assertion reads
$\delta_{\eps,0}^{\smallcap{h}}\cO_1 = F_1(\eps) - \cH(A_1(\eps)) - \vec{\cH}(\vec{B}_1(\eps))$,
where $F_1(\eps) = \{ \cV(\eps), \cO_0\}$ plays the role of $F$ in the Lemma.
In order to evaluate the integral term on the right-hand side of (\ref{OFvar}) we
note  $\delta_{\eps,0}^{\smallcap{h}}F_1(\alpha) -
\delta_{\alpha,0}^{\smallcap{h}}F_1(\eps) =
\{ \delta_{\eps,0}^{\smallcap{h}} \cV(\alpha) - \delta_{\alpha,0}^{\smallcap{h}} \cV(\eps), \cO_0\}$. It can be simplified using 
\be
\label{Vvarcom}
\delta_{\eps,0}^{\smallcap{h}}\cV(\alpha) - \delta_{\alpha,0}^{\smallcap{h}}\cV(\eps) =
\vec{\cH}( \vec{\gamma}(\alpha,\eps))\,,
\quad \gamma^a(\alpha,\eps) = q q^{ab} (\alpha \dd_b \eps - \eps \dd_b \alpha)\,,
\ee
This can be verified either by direct computation or by matching terms of
order $\check{\kappa}$ in the constraint algebra of Einstein gravity. So far,
\be
\label{O1check1} 
\delta_{\eps,0}^{\smallcap{h}}F_1(\alpha) - \delta_{\alpha,0}^{\smallcap{h}}F_1(\eps) =
\vec{\cH}(\{\vec{\gamma}(\alpha, \eps), \cO_0\})\,.
\ee
Upon evaluation on the interpolating flow we use
\be
\label{O1check2} 
\vec{\cH}|_s(\vec{\xi})
= \vec{\cH}(\vec{\xi}) + (s-1) \cH(\cL_{\vec{\xi}} T)\,,
\ee
where $\cL_{\vec{\xi}} T = \xi^a \dd_a T - \dd_a \xi^a T $ is the Lie derivative of
a spatial $-1$ density. This follows from the $d/ds$ differentiated version, which
in turn reflects the constraint algebra of VDC gravity,
i.e.~$\{ \cH(\eps_1), \cH(\eps_2)\} =0$, $\{ \vec{\cH}(\vec{\xi}), \cH(\eps)\}
= \cH(\cL_{\vec{\xi}}\,\eps)$. Combining (\ref{O1check1}) with (\ref{O1check2})
the integral term in the Lemma evaluates to 
\be
\label{ABeqn}
\int_0^1 \! ds s \big[ \delta_{\eps,0}^{\smallcap{h}}
  F_1(\alpha) - \delta_{\alpha,0}^{\smallcap{h}} F_1(\eps)
 \big]\big|_{s, \alpha =T} =  - \cH(A_1(\eps)) - \vec{\cH}(\vec{B}_1(\eps))\,,
\ee
with $A_1(\eps), \vec{B}_1(\eps)$ as in (\ref{AB1}). In summary,
(\ref{O1}) indeed solves the $j=1$ instance of (\ref{obsrecursion}), with
the specific $A_1(\eps), \vec{B}_1(\eps)$ given. 

Finally, we check that $\cO_1$ is a local functional on phase space,
specifically one of the form asserted in (\ref{Ojloc}), for $j=1$.  Mostly,
this follows from the fact that $\cV(\alpha)$ has regular
functional derivatives given by
\ba
\label{Vvar0}
\frac{\delta \cV(\alpha)}{\delta q_{ab}} \is
\alpha V^{ab} -
\frac{q}{2\kappa} (\nabla^a \nabla^b - q^{ab} \nabla^c \nabla_c)\alpha\,, 
\nonum
\frac{\delta \cV(\alpha)}{\delta \varphi} \is
-\frac{1}{\kappa} q q^{ab} \nabla_b (\alpha \nabla_a \vp) \,,
\ea
where we defined
\be
V^{ab} := \frac{q}{2\kappa} (R^{ab} - q^{ab} R) 
- \frac{q}{2\kappa} (q^{ad} q^{bc} - q^{ab} q^{cd}) \dd_c \vp \dd_d \vp\,.
\ee
By (\ref{O1}) we need to evaluate $\{\cV(\alpha), \cO_0\}$.  
Since $\cO_0 = \int\! d^3 x \,F_0(x)$ for some ultralocal $F_0$ its
functional derivatives reduce to pointwise ones, $\delta \cO_0/\delta q_{ab}(x) =
\dd F_0(x)/\dd q_{ab}(x)$, etc.. Further, as $\cV$ does not depend on
the momenta, the functional derivatives of $\cV$ with respect to the momenta vanish. Together 
\be
\label{G0} 
\{ \cV(\alpha), \cO_0\} = \int\! d^3 x \bigg( \frac{\delta \cV(\alpha)}{\delta q_{ab}(x)}
\frac{ \dd F_0(x)}{\dd p^{ab}(x)} +\frac{\delta \cV(\alpha)}{\delta \vp(x)}
\frac{ \dd F_0(x)}{\dd p^{\vp}(x)} \bigg)\,,
\ee
is a differential polynomial of order $2$ in the phase space variables. Since
$T(x)$ is an ultralocal $-1$ density on phase space, evaluation of
(\ref{G0}) on $\alpha =T$ will not change this.  Finally, the interpolating flow
$z^{(s)}$  arises from the pointwise in $x \in \Sigma$ integrated characteristic
flow $\dd_{\tau} z^f = \{ z^f, \cH\}$.  Hence
\be
\label{G1} 
G_1(x) = \int_0^1 \! ds \bigg(\frac{\delta \cV(\alpha)}{\delta q_{ab}(x)}
\frac{ \dd F_0(x)}{\dd p^{ab}(x)} +\frac{\delta \cV(\alpha)}{\delta \vp(x)}
\frac{ \dd F_0(x)}{\dd p^{\vp}(x)} \bigg)\bigg|_{s, \alpha = T}\,,
\ee
is a spatial differential polynomial in the phase space variables of degree $2$.   

The generalization to higher orders is considerably more complicated,
but the above presentation of the $j=1$ case gives a good flavor of the
reasoning involved. 



\newpage 
\begin{thebibliography}{99}
\begin{small}
  
\bibitem{ASvsCDT} J.~Ambjorn, J.~Gizbert-Studnicki, A.~Goerlich,
  and D.~Nemeth,
  Is lattice quantum gravity asymptotically safe? Making contact between causal dynamical triangulations and the functional renormalization group,
  Phys.\ Rev.\ {\bf D110} (2024) 126006. 

\bibitem{DonnGidd1} W.~Donnelly and S.~Giddings,
  Observables, gravitational dressing, and obstructions to
  locality and subsystems,
  Phys.\ Rev.\ {\bf D94} (2016) 104038.  
 
\bibitem{Constrdisc} B.~Dittrich and P.~H\"{o}hn,
  From covariant to canonical formulations of discrete gravity,
 Class.\ Quant.\ Grav.\ {\bf 27} (2010) 155001. 


\bibitem{ATorre1} I.~Anderson and C.~Torre,
Symmetries of the Einstein equations,
  Phys.\ Rev.\ Lett.\ {\bf 70} (1993) 3525. 

\bibitem{ATorre2} I.~Anderson and C.~Torre,
  Classification of local generalized symmetries for the vacuum Einstein equations,
  Commun.\ Math.\ Phys.\ {\bf 176} (1996) 479. 

\bibitem{Marchildon} L.~Marchildon,
Lie symmetries of Einstein's vacuum equations in $N$ dimensions,
  J.\ Nonlin.\ Math.\ Phys.\ {\bf 5}
(1998) 68.

\bibitem{TorreNoGo} C.~Torre,
Gravitational observables and local symmetries,
Phys.\ Rev.\ {\bf D48} (1993) R2373.

\bibitem{KucharNoGo}  K.~Kuchar,
 General relativity: dynamics without symmetry,
  J.\ Math.\ Phys.\ {\bf 22} (1981) 2640.
  
\bibitem{RingstroemCR}  H.~Ringstr\"{o}m, 
Cosmology, the Big Bang, and the BKL conjecture,
Comptes Rendus. M\'{e}canique, {\bf 353} (2025) 53.   

\bibitem{SvsCtensor} M.~Niedermaier,
Nonstandard action of diffeomorphisms and gravity's anti-Newtonian limit,
  Symmetry {\bf 12} (2020) 752.  

  
 


\bibitem{KhavkineObs} I.~Khavkine,
Local and gauge invariant observables in gravity,
  Class.\ Quant.\ Grav.\ {\bf 32} (2015) 185019.


\bibitem{Hoehn} C.~Goeller, P.~H\"{o}hn, and J.~Kirlin,
  Diffeomorphism-invariant observables and dynamical frames
  in gravity: reconciling bulk locality with general covariance,
  {\tt arXiv:2206.01193}. 

 \bibitem{DittTamb} B.~Dittrich and J.~Tambornino,
  A perturbative approach to Dirac observables and their spacetime algebra,
  Class.\ Quant.\ Grav.\ {\bf 24} (2007) 757.

\bibitem{Pons} J.~Pons, D.~Salisbury, and K.~Sundermeyer, 
Revisiting observables in generally covariant theories in the light of gauge fixing methods,
  Phys.\ Rev.\ {\bf D80} (2009) 084015. 


\bibitem{HTbook} M.~Henneaux and C.~Teitelboim, {\it Quantization of gauge
  systems}, Princeton UP, 1992. 

\bibitem{Ellisbook} G.~Ellis, R.~Maartens, M.~MacCallum,
  {\it Relativistic Cosmology}, Cambridge UP, 2012. 

 
\bibitem{Isham} C.~Isham,
Some quantum field theory aspects of the superspace quantization of general relativity,
  Proc.\ R.\ Soc.\ {\bf A351} (1976) 209. 
  
\bibitem{Henn1} M.~Henneaux,
Geometry of zero signature spacetimes,
Bull.\ Soc.\ Math.\ Belg.\ {\bf 31}, 47 (1979). 

\bibitem{Carrollgrav1}
  M. Henneaux and P. Salgado-Rebolledo,
Carroll contractions of Lorentz invariant theories,
  JHEP {\bf 11} (2021) 180.

\bibitem{Carrollgrav2} S.~Pekar, A.~Perez, and
  P.~Salgado-Rebolledo,
  Cartan-like formulation of electric Carrollian gravity,
  JHEP {\bf 09} (2024) 059.

\bibitem{Hartong} J.~Hartong,
 Gauging the Carroll algebra and ultrarelativistic gravity,
 JHEP {\bf 08} (2015) 69.

\bibitem{Primer} E.~Bergshoeff, J.~Figueroa-O'Farrill, and J.~Gomis,
  A non-lorentzian primer,
  SciPost Phys.\ Lect.\ Notes {\bf 69} (2023).

 \bibitem{CarrollHolo} L.~Donnay, A.~Fiorucci, Y.~Herfray, and
  R.~Ruzziconi,
  Carrollian perspective on Celestial Holography,
  Phys.\ Rev.\ Lett. {\bf 129} (2022) 071602.

  \bibitem{CarrollHoloRept} R.~Ruzziconi,
  Carrollian Physics and Holography,
  Phys.\ Rept.\ {\bf 1182} (2026) 1-87. 


\bibitem{CarrollGeom} L.~Ciambelli, R.~Leigh, C.~Marteau,
  and M.~Petropoulos,
  Carroll structures, Null geometry, and conformal isometries,
  Phys.\ Rev.\ {\bf D100} (2019) 046010. 
 
\bibitem{GRstrongI} M.~Niedermaier,
  The gauge structure of strong coupling gravity,
  Class.\ Quant.\ Grav.\ {\bf 32} (2015) 015007. 


  

\bibitem{Froeb} M.~Fr\"{o}b, C.~Rein, and R.~Verch,
  Graviton corrections to the Newtonian potential using invariant observables,
  JHEP {\bf 01} (2022) 180. 

  
\bibitem{GiddPerk} S.~Giddings and J.~Perkins,
  Perturbative quantum
  evolution of the gravitational state and dressing in general backgrounds,
  Phys.\ Rev.\ {\bf D110} (2024) 026012. 

\bibitem{Cheung} C.~Cheung, A.~Sivaramakrishnan, J.~Wilson-Gerow,
  and Lihang Zhou, Perturbatively dressed observables, Phys.\ Rev.\ {\bf D114} (2026) 086013. 
  

  \bibitem{PNresum1} P.~Damgaard and K.~Lee, Schwarzschild Black Hole from Perturbation Theory to all orders,
  Phys.\ Rev.\ Lett.\ {\bf 132} (2024) 251603.

\bibitem{PNresum2} M.~Saavedra,
  The classical gravitational impulse at high energies,
  {\tt arXiv:2605.18544}.



\bibitem{Gowdyobs} M.~Niedermaier and M.~Sedighi Jafari,
 Dirac observables for Gowdy cosmologies regular at the Big Bang, 
 SIGMA {\bf 22} (2026)  097. 

  
  \bibitem{Nikolaev} A.~Nikolaev,
 Kato perturbative expansion in classical mechanics and an
  explicit expression for the Deprit generator, 
  Theor.\ Math.\ Phys. {\bf 182} (2015) 407. 

  
\bibitem{Trivgrad} M.~Niedermaier,
  Canonical Trivialization of gravitational gradients, 
  Class.\ Quant.\ Grav.\ {\bf 34} (2017) 115013. 

\bibitem{Carrollexp1} D.~Hansen, N.~Obers, G.~Oling, and B.~Sogaard,
  Carroll expansion of general relativity,
  SciPost Phys.\ {\bf 13} (2022) 055.
  
 \end{small}
\end{thebibliography}
\end{document}